\documentclass[journal=jacsat,manuscript=article]{achemso}

\usepackage{multicol}
\usepackage[version=3]{mhchem} 

 \usepackage{mathrsfs}
 \usepackage{amsmath,amsfonts}
 \usepackage[normalem]{ulem} 
 \makeatletter
\newcommand*{\addFileDependency}[1]{
  \typeout{(#1)}
  \@addtofilelist{#1}
  \IfFileExists{#1}{}{\typeout{No file #1.}}
}
\makeatother

\usepackage{xr-hyper} 
\usepackage{hyperref}

\usepackage{graphicx}
\usepackage[T1]{fontenc}
\usepackage{cancel}
\usepackage{dirtytalk}
\usepackage{soul}
\usepackage{color}
\usepackage{float}
\newcommand{\cd}{$\text{Cd}_{\text{3}}\text{As}_{\text{2}}$}
\newcommand{\linb}{$\text{LiNbO}_{\text{3}}$}

\newcommand{\ttt}{\mathrm}
\newcommand{\te}[1]{$\mathrm{#1}$}

\usepackage{mathtools}

\DeclarePairedDelimiterX\braket[2]{\langle}{\rangle}{#1 \delimsize\vert #2}
\DeclarePairedDelimiter\abs{\lvert}{\rvert}%

\usepackage{ulem}
\usepackage{titlesec}
 \titleformat{\paragraph}[hang]{\bfseries}{}{0pt}{\uline}
\titlespacing*{\paragraph}
{0pt}{3.25ex plus 1ex minus .2ex}{1.5ex plus .2ex}

\newcommand{\energy}{\mathcal{E}}
\newcommand{\w}{\omega}
\newcommand{\bs}{\mathbf}

\newcommand{\lb}{\left(}
\newcommand{\rb}{\right)}
\newcommand{\ls}{\left[}
\newcommand{\rs}{\right]}
\newcommand{\lc}{\left\{}
\newcommand{\rc}{\right\}}
\author{Lu Wang}
\affiliation{School of Electrical and Electronic Engineering, Nanyang Technological University, 50 Nanyang Avenue, Singapore 639798, Singapore}
\author{Jeremy Lim}
\affiliation{Science, Math and Technology, Singapore University of Technology and Design, 8 Somapah Road, Singapore 487372, Singapore}
\author{Liang Jie Wong}
\affiliation{School of Electrical and Electronic Engineering, Nanyang Technological University, 50 Nanyang Avenue, Singapore 639798, Singapore}
\email{liangjie.wong@ntu.edu.sg}

\title[]
  {Highly efficient terahertz generation using 3D Dirac semimetals}

\abbreviations{THz}

\begin{document}


\begin{abstract}
We show that 3D Dirac semimetals are promising candidates for highly efficient optical-to-terahertz conversion due to their extreme optical nonlinearities. In particular, we predict that the conversion efficiency of \cd~exceeds typical materials like \linb~by $>5000$ times over nanoscale propagation distances. Our studies show that even when no restrictions are placed on propagation distance, \cd~still outperforms \linb~in efficiency by $>10$ times. Our results indicate that by tuning the Fermi energy, Pauli blocking can be leveraged to realize a step-like efficiency increase in the optical-to-terahertz conversion process. We find that large optical to terahertz conversion efficiencies persists over a wide range of input frequencies, input field strengths, Fermi energies, and temperatures. Our results could pave the way to the development of ultrathin-film terahertz sources for compact terahertz technologies.
\end{abstract}


\section{Introduction}
High energy, single-cycle terahertz pulses are essential for both fundamental studies and applications, such as materials analysis \cite{hamm2017perspective,manceau2010direct,sharma2010time}, 6G communications \cite{yang20196g}, electron acceleration \cite{zhang2018segmented}, and high resolution spectroscopy \cite{cocker2013ultrafast}. A common approach to realize high-energy, few-cycle terahertz pulses from optical pulses is optical rectification, which exploits the second-order nonlinearity of materials such as \linb~(LN), GaAs, ZnTe, GaP etc. \cite{fulop2020laser}. In particular, LN is widely used to generate terahertz pulses in the range of $1~\ttt{THz}$ due to its relatively strong nonlinearity \cite{hebling2004tunable}. Alternative platforms, such as graphene and gas plasma, have been studied for few-cycle terahertz pulse generation \cite{mikhailov2012theory,koulouklidis2020observation,sun2010coherent}. In all cases, strong nonlinearity is a key requirement for generating terahertz pulses of high energy.

Three-dimensional Dirac semimetals (3D DSMs) \cite{borisenko2014experimental}-- a recently discovered class of topological materials-- have been shown to exhibit extremely large optical nonlinearities that originate from their linear and gapless energy-momentum dispersion in all three dimensions. For this reason, the 3D DSM \cd~\cite{liu2014stable}, which possesses exceptionally high Fermi velocities and electron mobilities \cite{borisenko2014experimental,liu2014stable}, has been used to generate highly efficient terahertz high-order harmonics from input terahertz pulses \cite{cheng2020efficient,kovalev2020non,lim2020efficient,ullah2020third}, and studied as a platform for nonlinear plasmonics \cite{ooi2019nonlinear,ooi2020dirac}. Related materials like Weyl semimetals have been shown to support chiral terahertz emission and polarization manipulation \cite{gao2020chiral}.

We show that the extreme optical nonlinearity of 3D DSMs can be leveraged for highly efficient optical-to-terahertz conversion over nanometer-scale propagation distances. Specifically, we predict a conversion efficiency enhancement of over 5000 times in \cd~compared to LN, in a propagation distance of 300 \te{nm}. This is especially surprising given that we use the third-order nonlinearity in \cd, whereas the second-order nonlinearity is used for the corresponding process in LN. Our results reveal that tuning the Fermi energy allows us to leverage Pauli blocking to achieve a step-like conversion efficiency enhancement in terahertz generation. Our findings are crucial in the development of efficient ultrathin-film terahertz sources and the development of compact terahertz driven technologies \cite{withayachumnankul2012sub,lu2020strong}.
\begin{figure}[H]
\centering
   \includegraphics[width=1\linewidth]{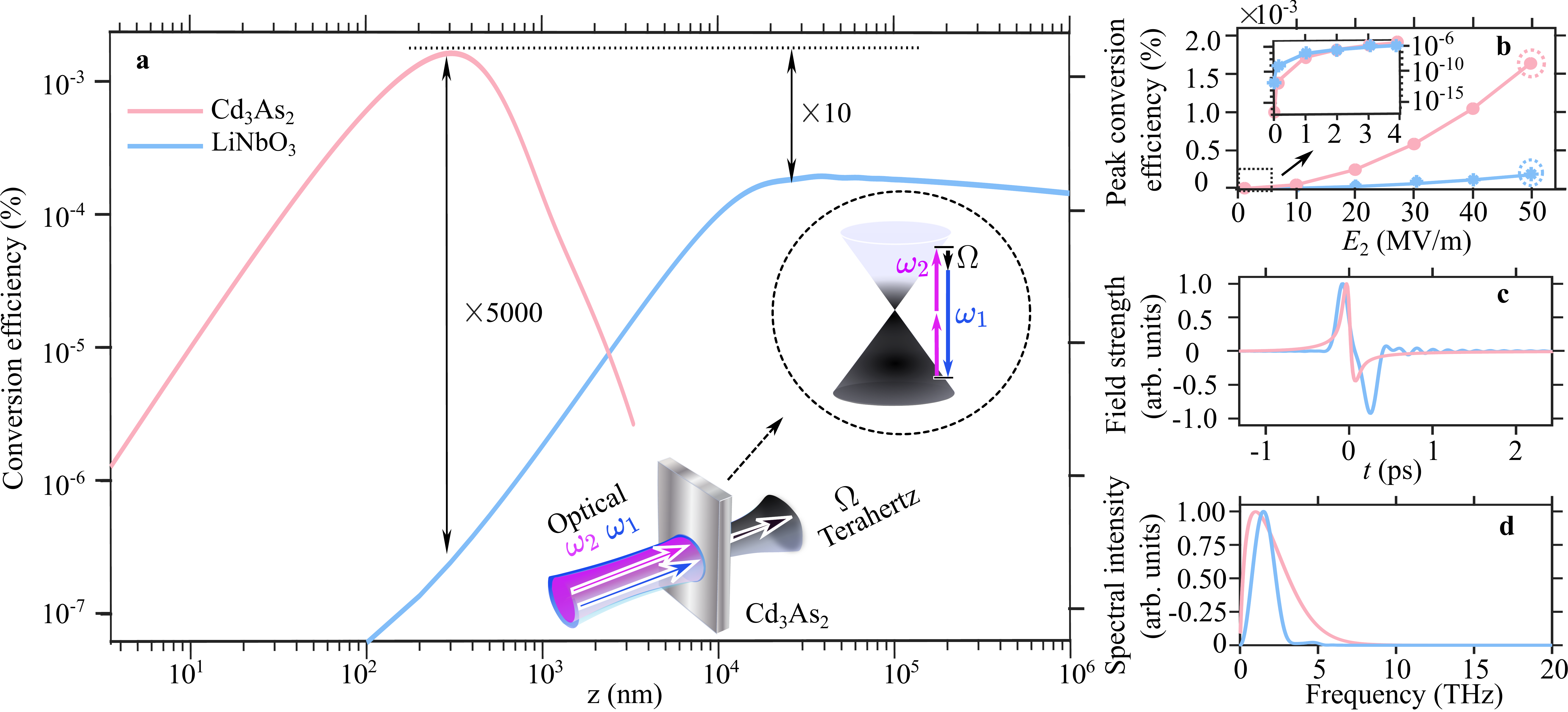}
\caption{Highly efficient terahertz generation via third-order nonlinearities in 3D DSMs. \textbf{a} Terahertz conversion efficiencies as a function of the propagation distance for an input field strength $E_2=50~\ttt{MV/m}$ (marked by dotted circles in \textbf{b}). The decrease of the efficiency is subjective to both the phase-matching condition and the material absorption. 3D DSM \cd~delivers over 5000 times the efficiency of \linb~(LN) over a propagation distance of 300 nm. \textbf{b} Terahertz peak conversion efficiency as a function of the input electric field strength $E_2$ centered at $\w_{20}$ (dots; curves are visual guides). The output terahertz electric fields and spectra at $E_2=50~\ttt{MV/m}$ (pink dotted circle in \textbf{b}) are presented in \textbf{c} and \textbf{d} respectively. We consider $\lambda_1=1~\ttt{\mu m}$, $\omega_{10}=2\pi c/\lambda_1$, and $\hbar\omega_{20}=\hbar\omega_{10}/2=0.62~\text{eV}$ at temperature $T=77~\ttt{K}$. We fix the Fermi energy at $\energy_\ttt{f}=0.45~\mathrm{eV}$ and scattering times for both the inter-band and intra-band to 150 \te{fs}. The electric field strength at $\w_1$ is fixed at $E_1=5~\mathrm{MV/m}$. Both the optical pulses have $150~\ttt{fs}$ full-width-half-maximum pulse duration. Unless otherwise stated, we consider these parameters throughout our work.}
\label{fig}
\end{figure}
\section{Model}
\subsection{Physics of terahertz generation in 3D DSMs}
Terahertz generation from optical pulses in DSMs, schematically illustrated in Fig. \ref{fig}\textbf{a} (inset), occurs when two co-propagating optical pulses of the same polarization ($\mathbf{\hat{x}}$  polarized), but with different central frequencies $\w_{10}$ and $\w_{20}$, (frequencies within the optical pulse centered at $\w_{10}$ and $\w_{20}$ are denoted by $\w_{1}$ and $\w_{2}$ respectively), impinge on the sample, generating an output terahertz pulse that travels in the same direction. In momentum space, the driving fields induce inter-band and intra-band carrier transitions, resulting in the absorption of two low-energy photons at $\omega_{2}\approx0.5\omega_1$ and emitting one high energy photon at $\omega_1$ and another low-energy terahertz photon of frequency $\Omega=2\omega_2-\omega_1$. In Fig. \ref{fig}, we study optical-to-terahertz conversion for the specific case of the 3D DSM \cd. We determine the linear and nonlinear material conductivities associated with the 3D Dirac cone band structure using perturbative quantum theory, and simulate the terahertz generation process by solving Maxwell's equations using these conducitvities. Our model fully considers effects including the optical Kerr effect, finite temperatures, arbitrary Fermi energies, and carrier scattering. Our model captures both inter-band and intra-band dynamics, as well as coupling between them (henceforth denoted by the term inter-intra-band). {The Hamiltonian that describes the carrier dynamics within a 3D DSM is given by}
\begin{equation}
{i} \hbar \partial_t \psi(t)= \left[H_0+H_\text{int} (t) \right]\psi(t)\label{ham},
\end{equation}
{where $\psi(t)$ is the electron wave function, $H_0= \sum v_{j} \sigma_{j} {p}_{j}$} is the stationary Hamiltonian, $H_\text{int}= e \bs{r} \cdot \bs{E}(t) $ is the interaction Hamiltonian in the length gauge \cite{aversa1995nonlinear}, $\bs{r}$ is the position operator, $e\,(>0)$ is the elementary charge, $\bs{E}$ is the electric field, $\hbar$ is the Planck constant, $\sigma_{j}$ is the Pauli matrix with ${j} \in {x,y,z}$, $v_{j}$ is the Fermi velocity along direction ${j}$ in Cartesian coordinates, and ${p_\ttt{j}}$ is the momentum operator in the ${j}$ direction. 

The length gauge is chosen for $H_\ttt{int}$ since the resulting nonlinear response is free of nonphysical zero-frequency divergences and a more transparent representation can be obtained \cite{aversa1995nonlinear,taghizadeh2017linear}. Due to the inversion symmetry of the \cd, even-ordered nonlinear conductivities are zero in our configuration. We apply perturbative quantum theory to Eq. (\ref{ham}) and obtain, for the first time, linear and nonlinear conductivities corresponding to a general 3D DSM (See Methods and SM Sections III and IV). Our model fully considers finite temperatures, carrier scattering, and an anisotropic Dirac cone band structures with Fermi velocities corresponding to realistic 3D DSM materials.

We simulate the terahertz generation process by solving the Maxwell's equations using a finite-difference split-step method, which captures linear and nonlinear propagation effects of paraxial pulses up to the third order in nonlinear conductivity, including the back conversion of the terahertz pulse on the optical pulses and the optical Kerr effect. In particular, by defining the electric field as
\begin{align}
& \tilde{\bs{E}}(z,t)=E\exp{(-t^2/\tau^2)}\exp{(-i\omega_0t)}\bs{\hat{x}}+\ttt{c.c}., \label{et} \\
& \mathscr{F}[ E\exp{(-t^2/\tau^2)}\exp{(-i\omega_0t)}]=E(z,\omega)\exp{[ik(\omega)z]}, \label{ew}
\end{align}
the optical-to-terahertz conversion in DSMs is given by  
\begin{eqnarray}
    {\partial E( z,\Omega)}/{\partial z}&&=\frac{-3}{2 n(\Omega)c \varepsilon_0}\int_0^{\infty}\!\!\int_0^{\infty} \!\!\!\! \sigma^{(3)}(\omega_2,\w_3,-\w_1)E_2(z,\omega_2) \nonumber\\
&&    \times E_1^*(z,\omega_2+\omega_3-\Omega) E_2(z,\omega_3) \exp \left\{ i \left[ k(\w_2)\right. \right.\nonumber\\
 && \left. \left. +k(\w_3)  -k^*(\w_1)-k(\Omega) \right] z\right\} d\omega_2 d\omega_3, \label{main_thz_gen}
\end{eqnarray}
where $\mathscr{F}$ is the Fourier transform, $^*$ represents complex conjugate, $z$ is the propagation distance, $\tau$ is the pulse duration, $\sigma^{(3)}(\w_2,\w_2,-\w_1)$ is the third-order conductivity, $\Omega$ represents the terahertz frequency,  $k(\w_1)=n(\w_1)\w_1/c$ represents the angular wavenumber at frequency $\w_1=\w_2+\w_3-\Omega$ (in our case both $\w_2$ and $\w_3$ are centered at $\w_{20}$), $n(\omega)$ is the refractive index, $z$ is the propagation distance, $c$ is the speed of light, and $\varepsilon_0$ is the vacuum permittivity. The terahertz pulse amplitude is given by $E$, and the optical pulse amplitudes by $E_1$ and $E_2$.
\begin{figure}[H]
\hspace*{-2cm}  
\centering
  \includegraphics[width=1.1\linewidth]{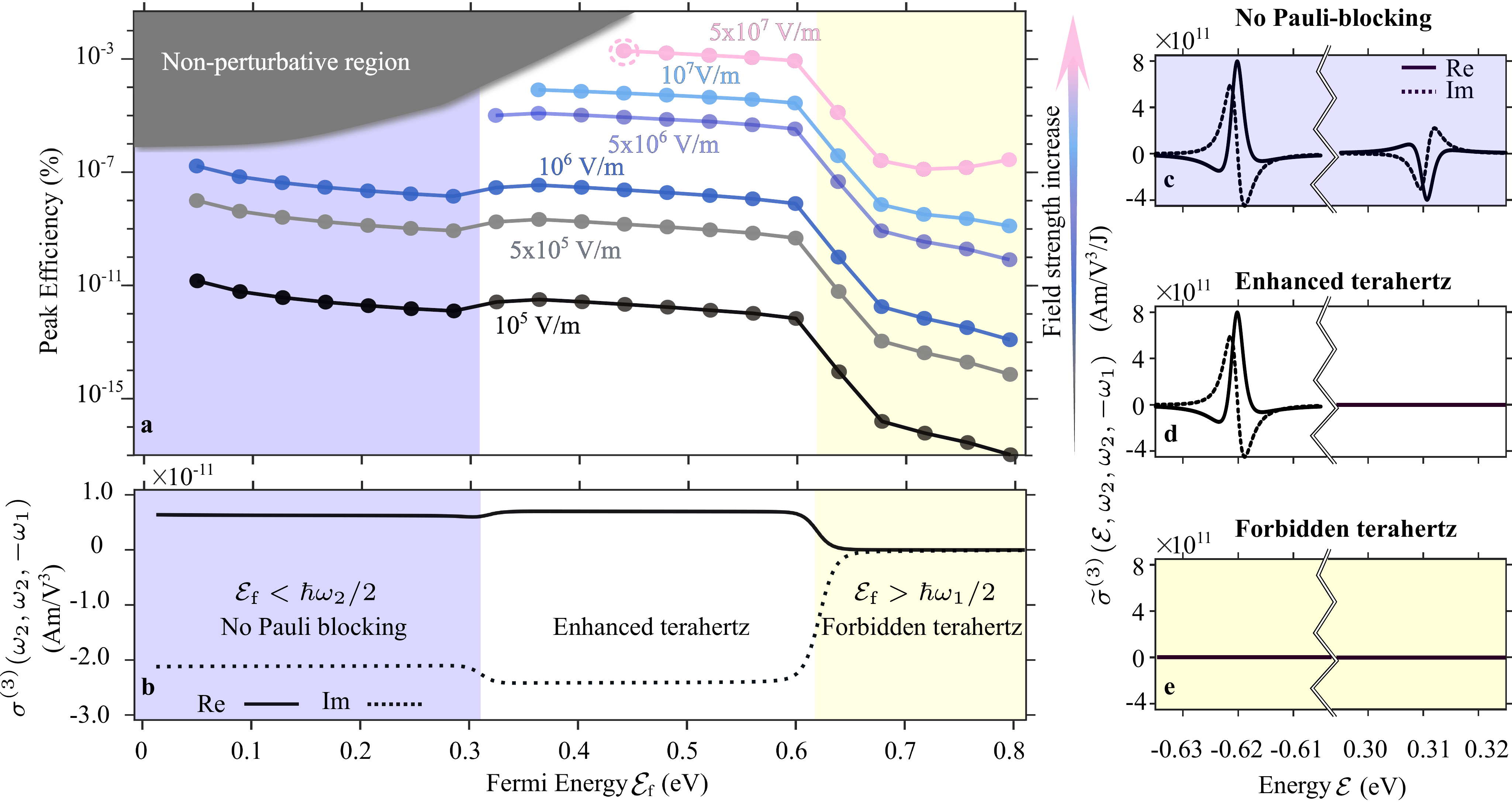}
\caption{Optical-to-terahertz conversion efficiency as a function of Fermi energy. \textbf{a} the peak terahertz conversion efficiencies at different input field strength $E_2$ as a function of the Fermi energy $\energy_\ttt{f}$. The simulation results are denoted by the filled circles. The curves are visual guides. The non-perturbative regime (gray shaded area) corresponds to $E_2<\sqrt{\abs{\sigma^{(1)}}/\abs{\sigma^{(3)}}}$ (see SM Section VII). The pink data point marked by the dashed circle indicates the same data as in Fig. \ref{fig}\textbf{b}. \textbf{b} shows how the third order conductivity for terahertz generation at 1 $\ttt{THz}$ ($\omega_{2}+\omega_{2}-\omega_{1}=1~\ttt{THz}$) varies as a function of $\energy_\ttt{f}$. The "No Pauli blocking", "{Enhanced terahertz}" and "{Forbidden terahertz}" regions in \textbf{b} correspond to \textbf{c} ($\energy_
\ttt{f}=0.05~\ttt{eV}$), \textbf{d} ($\energy_
\ttt{f}=0.45~\ttt{eV}$) and \textbf{e} ($\energy_
\ttt{f}=0.7~\ttt{eV}$) respectively, where the conductivity density {as a function of the electron energy} is shown (${\sigma}^{(3)}(\w_2,\w_2,-\w_1)=\int \widetilde{\sigma}^{(3)}(\energy,\w_2,\w_2,-\w_1) d\energy$).\textbf{c} shows two strong conductivity density peaks, which contribute with opposite signs to the overall conductivity. Consequently, an increase in the conductivity can be observed in \textbf{d} where one of the peaks are suppressed while the other remains. \textbf{e} shows that all conductivity density peaks are suppressed and thus the terahertz generation is forbidden. We consider the same simulation parameters as Fig. \ref{fig}. {It should be noted that linear and gapless band structure of \cd~ extends up to 1 eV \mbox{\cite{jeon2014landau,cheng2020efficient}}, which justifies plotting up to Fermi energies of 0.8 eV in (a) and (b). }}
\label{fig_fermi_scan}
\end{figure}
In this work, we consider the two optical pulses of amplitudes $E_1$ and $E_2$ centered at wavelengths of $1~\ttt{\mu m}~(\hbar\w_{10}=1.24~\ttt{eV})$ and $2~\ttt{\mu m}~(\hbar\w_{20}=0.62~\ttt{eV})$ respectively, {with $150~\ttt{fs}~(\tau=150~\ttt{fs}/\sqrt{2\log{(2)}})$ intensity full-width-half-maximum pulse duration. }
\section{Results}
\subsection{Enhanced optical optical-to-terahertz conversion efficiency in 3D DSMs}
Figure \ref{fig}\textbf{a} shows the optical-to-terahertz conversion efficiency as a function of propagation distance for a collinear configuration (Fig. \ref{fig}\textbf{a} inset). At a propagation distance of about 300 nm, the conversion efficiency in 3D DSM \cd~exceeds that of LN by $>$ 5000 times. Additionally, we find that even when no restrictions are placed on propagation distance, \cd~outperforms LN~in efficiency by over 10 times. Figure \ref{fig}\textbf{b} shows the terahertz conversion efficiency with respect to optical field $E_2$, when optical field $E_1$ is fixed at $5~\text{MV/m}$. Whereas LN~outperforms \cd~at low field strengths $E_2<\text{5~MV/m}$ (Fig. \ref{fig}\textbf{b}, inset), \cd~rapidly surpasses LN as field strength increases. The output terahertz fields and spectra for $E_2=50~\text{MV/m}$ (dotted circles in Fig. \ref{fig}\textbf{b}) are shown in Fig. \ref{fig}\textbf{c} and Fig. \ref{fig}\textbf{d} respectively. Here, we consider the experimentally measured Fermi velocities $(v_\ttt{x},v_\ttt{y},v_\ttt{z})=(1.28, 1.3, 0.33) \times 10^6 \ttt{m/s}$. It is possible that still higher conversion efficiencies exist at larger field strengths, but that would require a non-perturbative treatment for calculating the nonlinear conductivity that falls beyond the scope of this work. In SM Section VII, we present an analytical estimate for the input field strengths for which our conductivity calculations remain valid.
\subsection{Optimizing efficiency by tuning the Fermi energy }\label{fermi_sec}
Figure \ref{fig_fermi_scan} shows that an appropriate choice of the Fermi energy $\energy_\ttt{f}$ allows us to access a regime of enhanced terahertz generation. In Fig. \ref{fig_fermi_scan}\textbf{a}, we see that a broad range of Fermi energies and driving field strengths exist where substantial terahertz generation efficiencies can be accessed, even within the limits of perturbation theory.
{The generation efficiency is defined as the ratio to the generated terahertz energy and the input pump pulse energy, ${\int I(z,\Omega) d\Omega}/{\int I(0,\omega)d\omega}$,
where $I(z,\Omega)=c\epsilon_0 n(\omega)|E(z,\Omega)\exp{[ik(\Omega)z]}|^2/2$ is the intensity in the frequency domain as a function of terahertz frequency $\Omega$, $n(\omega)$ is the refractive index, and $\omega$ is the optical frequency.} The trend in conversion efficiency is partly explained through the third-order conductivity $\sigma^{(3)}(\omega_2,\omega_2,-\omega_1)$ in Fig. \ref{fig_fermi_scan}\textbf{b}, which follows a similar trend to the conversion efficiency as we increase the Fermi energy from the purple-shaded "No Pauli blocking" regime, to the unshaded "{Enhanced terahertz}" regime, and the yellow-shaded "{Forbidden terahertz}" regime. To further understand the step-like increase of the third-order conductivity, we plot in Figs. \ref{fig_fermi_scan}\textbf{c-d} the conductivity density
$\widetilde{\sigma}^{(3)}(\energy,\w_2,\w_2,-\w_1)$ as a function of the energy of the electronic states, defined by
\begin{equation}
    {\sigma}^{(3)}(\w_2,\w_2,-\w_1)=\int \widetilde{\sigma}^{(3)}(\energy,\w_2,\w_2,-\w_1) d\energy,
\end{equation}
at representative Fermi energy values from each regime ($0.05~\ttt{eV},~ 0.45~\ttt{eV},~0.7~\ttt{eV}$). {Here $\energy$ denotes the eigenenergy of an electron with a given wavevector $\bs{k}$}.

In Figs. \ref{fig_fermi_scan}\textbf{c}, we see that at relatively low Fermi energies, the nonlinear conductivity density corresponding to terahertz generation contains two peaks, one at $\energy \approx -0.62 ~\ttt{eV}$ and another at $\energy \approx 0.31 ~\ttt{eV}$. 
As the Fermi energy increases, the peaks of the conductivity density that lie within the range $\energy=[-\energy_\ttt{f}, \energy_\ttt{f}]$ are suppressed. We infer that this is related to Pauli blocking, which occurs when an electron cannot be excited from the valence band to the conduction band due to the lack of unoccupied states in the conduction band. This can be seen in Fig. \ref{fig_fermi_scan}\textbf{d}, where the peak at $\energy \approx 0.31 ~\ttt{eV}$ disappears. Since the contribution of the peaks at $\energy \approx -0.62 ~\ttt{eV}$ and $\energy \approx 0.31 ~\ttt{eV}$ add up destructively, the disappearance of one peak has the effect of enhancing the nonlinear conductivity associated with terahertz generation, explaining the step-like increase in conductivity moving from the "No Pauli blocking" to "Enhanced terahertz" regime. At still larger Fermi energies -- exemplified by the scenario in Fig. \ref{fig_fermi_scan}\textbf{e} -- both conductivity density peaks are suppressed since all transitions required for terahertz generation is forbidden, leading to a plunge in the resulting nonlinear conductivity in the "Forbidden terahertz" regime. 

In Fig. \ref{fig_fermi_scan}\textbf{a}, we see that the optical-to-terahertz conversion efficiency generally follows the same trend as the nonlinear conductivity in Fig. \ref{fig_fermi_scan}\textbf{b}. However, at low Fermi energies, the low electron filling in the conduction band leads to a smaller first-order conductivity at terahertz frequencies i.e. smaller terahertz absorption as shown in Eq. (\ref{intra_1}). Consequently, a relatively high efficiency can be potentially attained due to lower terahertz absorption in the "No Pauli blocking" regime at very low Fermi energies.

As Fig. 2b shows, the high conversion efficiency of \cd~ holds over a broad range of field strengths and Fermi energies. Although we have focused on the case of $T = 77 ~\ttt{K}$ here, our simulations at other temperatures (see Fig. 3 in SM Section V) reveal stability over a broad range of temperatures ranging from 4 K to 200 K. The enhanced conversion efficiency in the "Enhanced terahertz" regime, as well as the need to stay within the validity of our perturbative  conductivity calculations motivated our choice of $\energy_\ttt{f} = 0.45 ~eV$ in Fig. \ref{fig}. 
\section{Discussion}
\begin{figure}[H]
\centering
\includegraphics[width=0.6\linewidth]{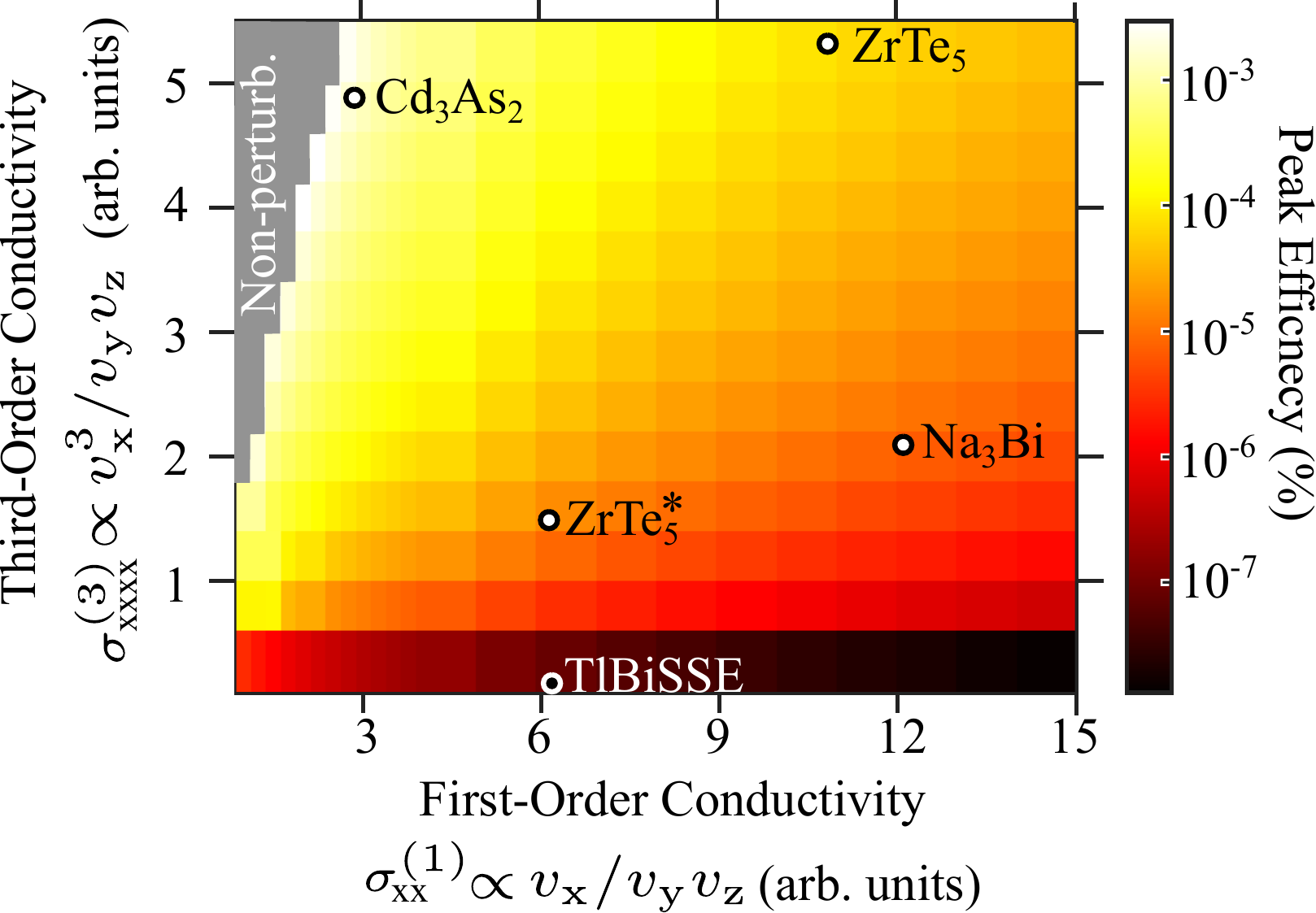}
\caption{Optical-to-terahertz conversion peak efficiencies efficiencies across the spectra of possible anisotropic 3D DSMs. The peak conversion efficiency is presented as a function of $\sigma^{(1)}_\text{xx} \propto v_\ttt{x}/(v_\ttt{y}v_\ttt{z})$ and $\sigma^{(3)}_\text{xxxx}\propto v_\ttt{x}^3/(v_\ttt{y}v_\ttt{z})$. The input electric field strengths are $E_1=5~\ttt{MV/m}$ and $E_2=50~\ttt{MV/m}$. The non-perturbative regime corresponds to the grey shaded area at the top-left corner. The above results are calculated with $\energy_\ttt{f}=0.45~\ttt{eV}$ at $T=77~\ttt{K}$. The Fermi velocities for the corresponding materials are listed in Table \ref{parameter}. We consider a propagation distance of 400 nm. Our results show that under the given conditions, \cd~is close to an ideal choice for optimal conversion efficiency.}
\label{fermi_v_raw}
\end{figure}

\begin{table}[h]
\begin{center}
\caption{Fermi velocities of different materials. The ZrTe$_5^*$ represents the averaged Fermi velocities $(v_\ttt{xy},v_\ttt{xz},v_\ttt{zy})$. For ZrTe$_5$, the Fermi velocity along $\hat{\bs{z}}$ is calculated by, $v_\ttt{z}=\sqrt{v_\ttt{xz}^2v_\ttt{zy}^2\text{\cite{zheng2016transport}}/(v_\ttt{x}v_\ttt{y} \text{\cite{martino2019two}})}$. }\label{parameter}
\begin{tabular}{ c |c }
\hline
Name & Fermi velocity $(v_\ttt{x},v_\ttt{y},v_\ttt{z})~\ttt{m/s}$ \\
\hline 
\cd & $(1.28, 1.3, 0.33) \times 10^6 $ \cite{liu2014stable}\\
Na$_3$Bi &$(4.17,3.63,0.95)\times 10^5 $ \cite{liu2014discovery}\\  
ZrTe$_5$  & $(7,4.6,1.94)\times 10^5$ \cite{zheng2016transport,martino2019two}  \\
ZrTe$_5^*$  & $(4.89,4.03,1.94)\times 10^5$ \cite{zheng2016transport}  \\
TlBiSSe & $(1.6,1.6,1.6)\times 10^5$ \cite{novak2015large}\\
 \hline
\end{tabular}
\end{center}
\end{table}
Our model is readily extended to capture the physics of a general, anisotropic 3D Dirac cone band structure. Although \cd~has been considered in Figs. \ref{fig} and \ref{fig_fermi_scan}, our model also applies to other 3D DSMs featuring different Fermi velocities. The significance of the material's Fermi velocities can be seen from Eqs. (\ref{intra_1}-\ref{sigma3_ei}) in Methods, where the linear conductivity $\sigma^{(1)}_\text{x}$ and the nonlinear conductivity $\sigma^{(3)}_\text{xxxx}$ are directly proportional to $v_\ttt{x}/(v_\ttt{y} v_\ttt{z})$ and $v_\ttt{x}^3/(v_\ttt{y} v_\ttt{z})$ respectively. Figure \ref{fermi_v_raw} shows the peak conversion efficiency as a function of these prefactors, revealing that the combination of a small $\sigma^{(1)}_\text{x}$ and a large $\sigma^{(3)}_\text{xxxx}$ can lead to efficient terahertz generation. This can also be understood intuitively since it implies low terahertz absorption and large nonlinearity for optical-to-terahertz conversion. We show that under the given conditions, \cd~is close to an ideal choice for optimal conversion efficiency.

Our findings suggest that even larger efficiencies can be obtained with field strengths and Fermi energies that require a non-perturbative treatment of the nonlinear conductivity. Larger conversion efficiencies for the same input fields could also potentially be obtained by considering a non-collinear interaction geometry. In particular, having obliquely incident input fields would lead to the existence of second-order nonlinearities in 3D DSMs, which could also be a promising avenue for efficient optical-to-terahertz conversion.

\section{Methods}
The input electric field and its positive frequency component $\exp{(-i\omega_0 t)}$ are shown in Eqs.~(\ref{et}) and (\ref{ew}) respectively. Without the loss of generality, the input fields are chosen to be linearly polarized in $\hat{\bs{x}}$ direction. Note that in the following calculations, only the positive frequency part of the $\sigma(\omega)$ and $E(\omega)$ are presented. However, no approximation is made. Since with the relation $\sigma(\omega)=\sigma(-\omega)^*$ \cite{sipe2000second} and $E(\omega)=E(-\omega)^*$, all the information is contained in the positive frequency elements.
The first-order conductivities are the following
\begin{eqnarray}
&&\sigma^{(1)}_{\text{i},\ttt{xx}}(\omega_1)=\frac{ig e^{2} v_{x}}{6 \pi^{2} \hbar^{3} v_{y} v_{z}(\omega_1+i\gamma_\ttt{i})} \left(\energy_\ttt{f}^{2}+\frac{\pi^{2}}{3} \ttt{k_{\mathrm{B}}}^{2} T^{2}\right), \label{intra_1}\\
&&\sigma^{(1)}_{\text{e},\ttt{xx}}(\omega_1)=\frac{i g e^2v_\ttt{x} (\w_1+i\gamma_\ttt{e})}{24v_\ttt{y}v_\ttt{z} \pi^2 \hbar}\int_{-\infty}^{\infty} \frac{n(\energy) }{\energy-\hbar(\omega_1+i\gamma_\ttt{e})/2} d\energy, \nonumber\\ \label{inter_1}
\end{eqnarray}
where "i" denotes intra-band, "e" denotes inter-band, $g=4$ is the combined valley and spin degeneracy, $n(\energy)=f(\energy)-f(-\energy)$, $f(\energy)=\lc\exp{\ls\lb\energy-\energy_\ttt{f}\rb/\ttt{k_B} T\rs}+1\rc^{-1}$ is the Fermi distribution, $\energy$ is the energy, $\ttt{k_B}$ is the Boltzmann constant, $T$ is the temperature, $\gamma_\ttt{i}$ is the inter-band decay rate, and $\gamma_\ttt{e}$ is the coherence decay rate (see SM). In our calculations, $\gamma_\ttt{e}=\gamma_\ttt{i}=1/(150~\ttt{fs})$.  Equation (\ref{inter_1}) agrees with the work of Kotov \cite{kotov2016dielectric} (see SM Eq. (39)). Equation (\ref{intra_1}) can also be obtained from the Boltzmann transport equation (SM Section VIII).

Although Eq.~(\ref{inter_1}) may appear to be readily solved via complex analysis, one should note that $n(\energy)$ contains an infinite number of poles in the complex-$\energy$ domain. By defining $\omega_\ttt{cv}=2\energy/\hbar$, the terms in the expression for the third order conductivity can be written as

\begin{align}
& \sigma^{(3)}_{\text{i},\ttt{xxxx}}(\w_1,\w_2,\w_{3})=\frac{ige^4v_\ttt{x}^3}{5\pi^2\hbar^3v_\ttt{y}v_\ttt{z}(\omega_1+\w_2+\w_3+i\gamma_\ttt{i})(\w_1+\w_2+i\gamma_\ttt{i})(\w_1+i\gamma_\ttt{i})} ,\label{sigma3iii}\\
&\sigma^{(3)}_{\text{e},\ttt{xxxx}}(\w_1,\w_2,\w_{3})=\frac{-ie^4v_\ttt{x}^3}{15\hbar^3\pi^2v_\ttt{y}v_\ttt{z}(\omega_1+\w_2+i\gamma_\ttt{i})}\int_{-\infty}^{\infty}\frac{2(\w_1+i\gamma_\ttt{e})n(\energy)/\energy}{ \w_\ttt{cv}^2-(\w_1+i\gamma_\ttt{e})^2}\nonumber\\
&\times \frac{1}{\w_\ttt{cv}-(\w_1+\w_2+\w_3+i\gamma_\ttt{e})}d\energy, \label{sigma3eee}\\
&\sigma^{(3)}_{\text{ie},\ttt{xxxx}}(\w_1,\w_2,\w_{3})=\frac{-ige^4v_\ttt{x}^3}{30v_\ttt{y}v_\ttt{z}\pi^2\hbar^3(\w_1+\w_2+\w_3+i\gamma_\ttt{i})}\int_{-\infty}^{\infty} \left\{ \frac{4 n(\energy)/\energy}{(\w_1+\w_2+i\gamma_\ttt{i})}  \frac{1}{(\w_\ttt{cv}-\w_1-i\gamma_\ttt{e})} \right.\nonumber\\
&\left. +\frac{\mathcal{M}_{-}(\w_1,\energy)}{\w_\ttt{cv}-\w_1-\w_2-i\gamma_\ttt{e}} \right\} d \energy, \\
&\sigma^{(3)}_{\ttt{ei},\ttt{xxxx}}(\w_1,\w_2,\w_{3})=\frac{-ige^4v_\ttt{x}^3}{30v_\ttt{y}v_\ttt{z}\pi^2\hbar^3}\left\{ \frac{1}{(\w_1+i\gamma_\ttt{i})(\w_1+\w_2+i\gamma_\ttt{i})}\int_{-\infty}^{\infty} \frac{4\partial_\energy n(\energy)+\energy \partial^2_\energy n(\energy)}{(\w_\ttt{cv}-\w_1-\w_2-\w_3-i\gamma_\ttt{e})} \right. \nonumber \\
& \left. +\int_{-\infty}^{\infty} \frac{(2\w_\ttt{cv}-\w_1-\w_2-\w_3-i\gamma_\ttt{e})\mathcal{M}_{-}(\w_1,\energy)}{(\w_\ttt{cv}-\w_1-\w_2-\w_3-i\gamma_\ttt{e})^2(\w_\ttt{cv}-\w_1-\w_2-i\gamma_\ttt{e})} \right\} d\energy, \label{sigma3_ei}\\
 &\mathcal{M}_{-}(\w_1,\energy)=\left[\frac{\partial}{\partial \energy}  \lb\frac{n}{\w_\ttt{cv}-\w_1-i\gamma_\ttt{e}}\rb  
 -\frac{2n}{(\w_\ttt{cv}-\w_1-i\gamma_\ttt{e})\energy}-\frac{\partial_\energy n(\energy)}{\w_1+i\gamma_\ttt{i}}\right],
\end{align}

where $\sigma^{(3)}_{\text{i},\ttt{xxxx}}$ represents the purely intra-band process, $\sigma^{(3)}_{\text{e},\ttt{xxxx}}$ represents the inter-band process, $\sigma^{(3)}_{\text{ie},\ttt{xxxx}}$ and $\sigma^{(3)}_{\text{ei},\ttt{xxxx}}$ arise from the diagonal terms of the density matrix and the coherence terms of the density matrix respectively and represent the inter-intra-band process (see SM). The frequency permutation remains to be included in Eqs.~(\ref{sigma3iii}-\ref{sigma3_ei}), due to the permutation symmetry \cite{boyd2020nonlinear}. The final $\sigma^{(n)}$ should be an average of all ($n!$) frequency permutations i.e. $\sigma^{(3)}=\mathcal{P}\ls \sigma^{(3)}_\text{i}+\sigma^{(3)}_\text{e}+\sigma^{(3)}_\text{ie}+\sigma^{(3)}_\text{ei}\rs/6$, where $\mathcal{P}$ represents the summation of 6 possible permutations. Similar as Eq. (\ref{intra_1}), Eq. (\ref{sigma3iii}) can also be obtained by Boltzmann transport equation (SM Section VIII). Surprisingly, $\sigma^{(3)}_\text{i}$ is independent of $\energy_\ttt{f}$ and $T$. The lack of dependence on $\energy_\ttt{f}$ for $\sigma^{(3)}_\text{i}$ has also been discussed in the work of Cheng et. al \cite{cheng2020third}, which present linear and nonlinear conductivity expressions for isotropic 3D DSMs in the limit where $T=0$ and carrier scattering is negligible.
\section{Conclusion}
In conclusion, we find that \cd~ is a promising platform for highly efficient terahertz generation. We predict an enhancement in efficiency of $>5000$ times in 3D DSMs \cd, compared to conventional materials like \linb, over a nanoscale propagation distance. Even when no restrictions are placed on propagation distance, \cd~still outperforms \linb~in efficiency by over $10$ times. 
Furthermore, our results indicate that tuning the Fermi energy allows us to leverage Pauli blocking, which can be leveraged to realize a step-like efficiency increase in the optical-to-terahertz conversion process. We achieve these exciting results despite working within the perturbative regime, and we expect similarly promising results at non-perturbative fields strengths, a regime which warrants future investigation. We also present closed form expressions for linear and nonlinear conductivities that take into account the effect of finite temperatures, carrier scattering, and anisotropic Dirac cone band structures with Fermi velocities corresponding to realistic 3D DSM materials. Our findings should pave the way towards the development of efficient ultrathin-film terahertz sources for compact terahertz driven technologies.

\begin{acknowledgement}
This work is funded by Singapore Institute of Manufacturing Technology (Singapore Institute of Manufacturing Technology - A STAR)-A1984c0043.

\end{acknowledgement}

\begin{suppinfo}

The data is available upon reasonable request.

\end{suppinfo}
\noindent Notes: The authors declare no competing financial interest.
\bibliography{achemso-demo}

\end{document}